1

**The unperformed experiment as an act of annihilation**

Thomas V Marcella[*]

Department of Physics and Applied Physics, University of Massachusetts at Lowell, Lowell, Massachusetts, 01854

Here, we accept the results of quantum experiments at face value and we make no apology for the failure of classical physics. Just as Wheeler has called particle detection an "act of creation", we suggest that, in some circumstances, not detecting the particle might be considered an "act of annihilation". We propose a delayed choice experiment in which we decide to interrupt an experiment after the particle has supposedly passed through the preparation apparatus and is on the verge of being detected. Even with the detection device in place, the particle is nowhere to be found.

PACS number(s): 03.65.Ta, 03.65.-w

I.     INTRODUCTION

Quantum mechanics is a statistical theory about observables and their measurement. It is a stingy theory, predicting only the possible results of a measurement and the probability distribution of those results. It does not pretend to describe any physical reality. Unlike classical physics, it does not explain, "how things happen". Yet, it correctly accounts for all the information that is revealed to us in a quantum event.

In this presentation, we accept the results of quantum experiments at face value and we do not speculate about those things not supported by experimental evidence. Nor do we embrace any particular interpretation [1]. We discuss the nature of the quantum experiment in light of evidence gleaned from actual experiments, especially delayed choice experiments [2,3,4,5] and experiments designed to test Bells theorem [6,7,8]. If experiment shows that nature sometimes violates the most fundamental tenets of classical physics, then, so be it.

Within this rigid framework we describe the intrinsic nature of a quantum experiment. We emphasize three non-classical characteristics of a quantum experiment that were first enunciated by Bohr [9]. First, the entire experiment including the detector and measurement result is a single entity. It is indivisible and it cannot be broken into its separate parts. Nor is it a sequence of physical events. The results obtained depend on the whole experimental arrangement.

Second, Bohr recognized that the experiment is not complete without a measurement result. The result is an irreversible event that gives closure to the experiment. As expressed by Wheeler, "No elementary phenomenon is a phenomenon until it is a registered phenomenon [3]."

Third, if we change the apparatus at any time during the experiment, the results obtained correspond to the experimental configuration in place at the moment the experiment is closed. Only the final configuration matters. Delayed choice experiments confirm such behavior.

In section III we propose a delayed choice experiment in which the final configuration is an unperformed experiment [10]. Other experiments indicate that closure occurs at the moment of particle detection. But, in the experiment described here, the particle is not detected. Consequently, there is no result and no closure.

## II.    THE QUANTUM EXPERIMENT

We construct an experiment from classical devices that are well known to us. Essentially, the experiment consists of a particle source, a preparation apparatus, a measuring device, and a measurement result. We imagine that the particle is emitted from the source and travels through the preparation apparatus and into the measurement device. The final step in this (classical) sequence of events is detection of the particle. The experiment is not complete until this occurs. Although the components are all classical objects there is no classical explanation of "how the experiment works".

All the different elements of the experimental apparatus make up a whole with no classical analog. The entire experimental apparatus, including the particle and the measurement result, is a single quantum entity, a phenomenon that cannot be subdivided into separate parts. Such a system does not satisfy the locality principle of classical physics.

The experimental result is a measured value of a specified observable. It is an essential part of the experiment and it gives finality to the experiment. Delayed choice experiments confirm that the measurement results, as well as the statistical distribution of those results, are determined by the experimental configuration in place at the instant the particle is detected, even when changes are made at the last possible moment, long after the particle has supposedly passed through the preparation apparatus. Thus, the experiment itself is not realized until it is closed. Paraphrasing Wheeler, no experiment is an experiment until there is an experimental result. Real experiments have results. Unperformed experiments have none.

The measurement result does not refer to an objective property associated with the particle as it supposedly makes its way through the apparatus; the result is not related to any preexisting conditions. Rather, the result corresponds to a particular experimental arrangement. It is a number randomly generated by the entire apparatus, including the detection device, at the instant the particle is detected. The quantum experiment is not a (classical) sequence of events leading up to particle detection, and the assumed track of the particle through the apparatus is illusory. As Wheeler has admonished us, " ------ we have no right to say what the photon is doing in all its long course from point of entry to point of detection."

Thus, the pre-measurement experiment, and post-measurement experiment as well, are undefined. They are unperformed experiments. There is no "before" or "after". Any preexisting conditions prior to measurement cannot be verified and are not revealed to us in the existing experiment. In particular, quantum theory is incompatible with the concept of preexisting properties of the particle. We have only the experimental configuration and the measurement result at the instant the particle is detected. In this sense, Wheeler calls the measurement process an "elementary act of creation".

## III     DELAYED CHOICE UNPERFORMED EXPERIMENT

An "unperformed experiment" refers to any result that is not realizable in the current experiment. A real experiment requires that the particle trigger a detector. If, for any reason, this does not occur, then there is no corresponding experiment. For example, consider an arrangement in which the particle is blocked somewhere on its way to the measuring device. Then, it is not available for detection and the experiment is not complete. In such a case, the entire apparatus, including the particle detector, constitutes an unperformed experiment.

We now propose a delayed choice experiment in which, at the very last moment, we reconfigure the experimental apparatus to be such an unperformed experiment. A simple example is particle diffraction at a single slit where the slit is equipped with an opaque shutter that can be inserted or removed at will. Without the shutter in place, the particle is detected and repeated measurements exhibit the usual diffraction pattern.

We assume a particle source that emits particles so slowly that there is only one particle in the apparatus at any time. We now insert the shutter sometime after the particle should have passed through the slit and is supposed to be somewhere between the slit and the detector. Now, the final experimental configuration, the one that defines the phenomenon, is an unperformed experiment. With the shutter in place, then, we "observe" nothing, which is the non-result of an unperformed experiment.

We repeat the experiment many times, always inserting the shutter at the last moment, but the detector is never triggered, no matter how long we wait. The unperformed experiment yields no information about the particle that was thought to exist when we constructed the experiment. The particle is nowhere to be found, even with the detector in place. A particle, apparently on the verge of detection, has suddenly disappeared! The unperformed experiment has become an "act of annihilation".

As Wheeler suggested, the particle exists only as a result of being detected. Here, we deny its existence by choosing the unperformed experiment. In other delayed choice experiments the particle is ultimately found when we look for it. Here, we look for the particle, but cannot find it.

## IV     CONCLUDING REMARKS

If we accept this annihilation experiment for what it is, then we have a trivial example of quantum mechanics at work. There is no measurement result and no quantum phenomenon to be observed. There is nothing more to discuss about this quantum non-event!

Many will not accept such a harsh, cursory description. They insist on asking, "What happened to the particle?" Unfortunately, there is no answer to that question. A quantum experiment reveals no evidence for the existence of any particle prior to measurement and we cannot verify that any particle actually vanished during the annihilation experiment.

Our discomfort is due to the failure of classical reasoning. We had tacitly done a classical time-of-flight calculation to determine when the particle had passed beyond the slits. Such a calculation is meaningless. In truth, we do not know where the particle was, assuming it was somewhere, when we put the shutter in place. Only if we insist on a classical explanation, do we encounter an unexplainable event.

As Bohr insisted, the particle is not distinct from the rest of the experimental apparatus and it should not be treated as such. Quantum mechanics describes the entire experiment and, in particular, it does not describe the local behavior of the particle. Thus, we should not expect this so-called annihilation experiment to do otherwise.